\documentclass[a4paper,11pt]{article}

\usepackage{amsmath,amssymb}     
\usepackage{color}
\usepackage{graphicx}
\usepackage{subfigure}
\usepackage{cite}                
\usepackage{hyperref}            
\usepackage{multirow,makecell}   
\usepackage{simplewick}
\numberwithin{equation}{section}   

\def \be {\begin{equation}}
\def \ee {\end{equation}}
\def \ba {\begin{array}}
\def \ea {\end{array}}
\def \bea{\begin{eqnarray}}
\def \eea{\end{eqnarray}}
\def \nn {\nonumber}

\def \ve {\varepsilon}
\def \m {\mu}
\def \n {\nu}

\def \r {\rho}

\def \p {\partial}
\def \f {\frac}

\def \ii {\mathrm{i}}

\def \tr {\textrm{tr}}

\begin{document}

\begin{titlepage}
	
	\title{\textbf {Open Spin Chains from Determinant Like Operators in ABJM Theory}}
	\author{Hui-Huang Chen$^{a}$\footnote{chenhh@jxnu.edu.cn}~,  Hao Ouyang$^{b}$\footnote{hao.ouyang@su.se}~,
		Jun-Bao Wu$^{c, d}$\footnote{junbao.wu@tju.edu.cn}}
	\date{}
\begin{flushright}\footnotesize
	
	\texttt{NORDITA-2018-090} \\
	
\end{flushright}	
{\let\newpage\relax\maketitle}
	\maketitle
	\underline{}
	\vspace{-10mm}
	
	\begin{center}
		{\it
			$^{a}$ College of Physics and Communication Electronics, Jiangxi Normal University, \\Nanchang 330022, China\\
			$^{b}$ Nordita, KTH Royal Institute of Technology and Stockholm University,
			Roslagstullsbacken 23, SE-106 91 Stockholm, Sweden\\
			$^{c}$ Center for Joint Quantum Studies and  Department of Physics, School of Science, Tianjin University, 135 Yaguan Road, Tianjin 300350, China\\
			$^{d}$ Center for High Energy Physics, Peking University, 5 Yiheyuan Road, Beijing 100871, China
		}
		\vspace{10mm}
	\end{center}
	\begin{abstract}
		We study the mixing problem of the determinant like operators in ABJM theory to two loop order in the scalar sector. The gravity duals of these  operators are open strings attached to the maximal giant graviton, which is a D4-brane wrapping a $\mathbb{CP}^2$ inside $\mathbb{CP}^3$ in our case. The anomalous dimension matrix of these operators can be regarded as an open spin chain Hamiltonian. We provide strong evidence of its integrability based on coordinate Bethe ansatz method and boundary Yang-Baxter equation.
	\end{abstract}
	
\end{titlepage}
\section{Introduction}

In recent years, a lot of progresses have been made in applying techniques of integrability to planar $\mathrm{AdS}_5/\mathrm{CFT}_4$ correspondence between IIB superstring theory on $\mathrm{AdS}_5 \times S^5$ and  four dimensional $\mathcal{N}=4$ super Yang-Mills (SYM) theory, see \cite{Beisert:2010jr} for a collection of reviews. Among all these notable progresses, spin chains or strings with periodic boundary condition are mostly studied and understood very well. People are also interested in non-periodic cases, including twisted boundary conditions, see for example \cite{Beisert:2005if,Beisert:2005he} and open boundary conditions\cite{Chen:2004,Chen:2004yf,DeWolfe:2004,Erler:2005nr,Berenstein:2005vf}. See \cite{Zoubos:2010kh,vanTongeren:2013gva} as  reviews for these interesting topics. \\
\par In 2008, another example of $\mathrm{AdS}/\mathrm{CFT}$ was proposed in \cite{Aharony:2008ug}, where the authors gave very strong evidence that type \uppercase\expandafter{\romannumeral2}A string theory on $\mathrm{AdS}_4\times\mathbb{CP}^3$ background is dual to $\mathcal{N}=6$ superconformal Chern-Simons
matter theory (also known as Aharony-Bergman-Jafferis-Maldacena(ABJM) theory) in three dimensional spacetime  with gauge group $U(N)\times U(N)$ and Chern-Simons levels $(k,-k)$. The 't Hooft
coupling of ABJM theory  turns out to be $\lambda=N/k$. People usually call this dual as $\mathrm{AdS}_4/\mathrm{CFT}_3$ correspondence or $\mathrm{ABJM}/\mathrm{AdS}_4\times\mathbb{CP}^3$ correspondence. Integrable structure in this setup was also extensively studied \cite{Klose:2010ki}. \\

\par Along similar path, many studies on non-periodic integrable cases re-emerged in the context of ABJM theory \cite{He:2013hxd,Chen:2016geo,Chen:2016niy,Bai:2016pxs,Bai:2017jpe}. However, there are still some potential integrable setups have not been investigated in the $\mathrm{AdS}_4/\mathrm{CFT}_3$ case, such as integrable Wilson loops \cite{Drukker:2006xg, Drukker:2012de, Correa:2012hh} and integrability from giant gravitons\cite{Berenstein:2005vf,Hofman:2007xp} found in the $\mathcal{N}=4$ SYM theory. In the SYM context, determinant like operators are dual to open strings attached to D-branes wrapping cycles in $S^5$. In the gravity side, such D-branes wrapping some cycles and carrying some angular momentum are usually called giant gravitons. In the context of $\mathcal{N}=4$ SYM, integrablity of open chain from giant gravitons have been studied extensively \cite{Berenstein:2005vf,Hofman:2007xp,Berenstein:2005fa,Ahn:2008df,Bajnok:2012xc,Bajnok:2013wsa}.
However  such integrable structure from the giant gravitons in the $\mathrm{AdS}_4/\mathrm{CFT}_3$  \cite{Berenstein:2008dc,Giovannoni:2011pn,Lozano:2013ota} case has not been explored as far as we know, though the the plane wave limit in both sides are studied in \cite{Cardona:2014ora}. In this paper, we would like to take a first step to fill these gaps. We study the anomalous dimension matrix of the determinant like scalar operators in ABJM theory up to two-loop order in the scalar sector. The anomalous dimension matrix can be viewed as the Hamiltonian of an open spin chain. Using the coordinate Bethe ansatz method, we  calculate the reflection matrix for fundamental excitations of this open chain. Based on the known bulk two body S-matrix, it is not hard to verify that the boundary Yang-Baxter equations (reflection equations) are  satisfied, hinting that this open spin chain is integrable.\\
\par The outline of this paper is as follows. In section 2, we introduce the determinant like scalar operators in ABJM theory. To study the mixing problem, we calculate their two point functions to two-loop order, giving the Hamiltonian of an open spin chain. In section 3, we compute the reflection matrix of this open spin chain through the coordinate Bethe ansatz method. Borrowing the two body S-matrix in the bulk from the previous result in \cite{Ahn:2009zg}, we confirm that the boundary Yang-Baxter equations (reflection equations) are satisfied. In the last section, we conclude and briefly discuss some possible problems  for further studies.

\section{Open spin chain in ABJM theory}
\subsection{Determinant like operators in ABJM theory}
  We begin with a very brief review of determinant like operators in ABJM theory.
 In ABJM theory, the scalar fields $(A_1, A_2, B_1^\dag, B_2^\dag)$ transform in the fundamental representation of the $SU(4)$ R-symmetry group. We make the following identification,
 \begin{equation}
 (A_1, A_2, B_1^\dag, B_2^\dag)=(Y_1, Y_2, Y_3, Y_4).
 \end{equation}
 Using  the conventions of \cite{Benna:2008zy}, the action of ABJM theory can be written as
\begin{equation}
\begin{split}
&S =\int d^3x(L_{CS} +L_k -V_F-V_B),\\
&L_{CS}= \f{k}{4\pi}\ve^{\m\n\r}\tr \Big( A_\m\p_\n A_\r +\f{2\ii}{3}A_\m A_\n A_\r
-\hat{A}_\m\p_\n \hat{A}_\r -\f{2\ii}{3}\hat{A}_\m \hat{ A}_\n \hat{A}_\r  \Big),\\
&L_k = \tr( -D_\m Y^\dag_I D^\m Y^I + \ii\Psi^{\dag I}\gamma^\m D_\m\Psi_I ),\\
&V_F =\dfrac{2\pi i}{k} \tr \Big( Y^\dag_I Y^I \Psi^{\dag J} \Psi_J
- 2 Y^\dag_I Y^J  \Psi^{\dag I} \Psi_J  + \epsilon^{IJKL} Y^\dag_I \Psi_J Y^\dag_K \Psi_L
 \\
&\phantom{L_Y =}-Y^I Y^\dag_I \Psi_J  \Psi^{\dag J}
- 2 Y^I Y^\dag_J \Psi_I \Psi^{\dag J} + \epsilon_{IJKL} Y^I \Psi^{\dag J} Y^K \Psi^{\dag L} \Big),
\\
&V_B=-\dfrac{4\pi^2}{3k^2}
\tr\Big(
Y_I^\dag Y^J Y_J^\dag Y^K Y^\dag_K Y^I+
Y_I^\dag Y^I Y_J^\dag Y^J Y^\dag_K Y^K+4
Y_I^\dag Y^J Y_K^\dag Y^I Y^\dag_J Y^K \\
&\phantom{L_Y =}
-6Y_I^\dag Y^I Y_J^\dag Y^K Y^\dag_K Y^J
\Big).
\end{split}
\end{equation}
Covariant derivatives are defined as
\begin{equation}
\begin{split}
&D_\mu Y^I=\partial_\mu Y^I+i A_\mu Y^I-i  Y^I \hat A_\mu,~~~D_\mu Y_I^\dag=\partial_\mu Y_I^\dag+i \hat A_\mu Y_I^\dag-i  Y_I^\dag A_\mu\\
&D_\mu \Psi_I=\partial_\mu \Psi_I+i A_\mu \Psi_I-i  \Psi_I \hat A_\mu.
\end{split}
\end{equation}

In this paper we focus on the determinant like operators
\begin{equation}\label{op1}
O_W = \epsilon_{a_1...a_N}\epsilon^{b_1...b_N}(A_1 B_1)^{a_1}_{b_1}...(A_1 B_1)^{a_{N-1}}_{b_{N-1}}W^{a_N}_{b_N},
\end{equation}
with
\begin{equation}\label{W}
W=Y^{I_1}Y_{J_1}^\dagger \cdots Y^{I_L}Y_{J_L}^\dagger.
\end{equation}
It was suggested in \cite{Cardona:2014ora} that the dual descriptions of these operators are open strings attached to the giant graviton D4-brane wrapping a $\mathbb{CP}^2$ inside $\mathbb{CP}^3$. The operator with $W=A_1B_1$ is dual to the D4-brane itself.

As discussed in \cite{Berenstein:2005vf}, open spin chain corresponding to determinant like operators in $\mathcal{N}=4$ SYM has nontrivial boundary conditions.
One may expect that there are similar boundary conditions in the case of open spin chain in  ABJM theory. To show this, we compute the tree level two-point function. The operator $O_W$ and its conjugate $\bar O_{W}$ can be rewrite as
\begin{equation}
\begin{split}
O_W = & \frac{1}{(N-1)!}\epsilon^{[I]_{N-1}c}_{[J]_{N-1}a}\epsilon^{[K]_{N-1}b}_{[L]_{N-1}c}
 A^{[J]_{N-1}}_{[I]_{N-1}}B^{[L]_{N-1}}_{[K]_{N-1}} W^a_b,\\
 \bar O_{W}  =& \frac{1}{(N-1)!}\epsilon^{[M]_{N-1}f}_{[S]_{N-1}d}\epsilon^{[Q]_{N-1}e}_{[P]_{N-1}f}
\bar A^{[P]_{N-1}}_{[Q]_{N-1}}\bar B^{[S]_{N-1}}_{[M]_{N-1}}\bar{W}^d_e.
\end{split}
\end{equation}
Here we use the shorthand notations
\begin{equation}
\begin{split}
&A=A_1,~~~B=B_1,~~~\bar A=A_1^\dag,~~~\bar B=B_1^\dag,\\
&[I]_{N-1}=I_1...I_{N-1},~~~A^{[J]_{N-1}}_{[I]_{N-1}}=A^{J_1}_{I_1}...A^{J_{N-1}}_{I_{N-1}}.
\end{split}
\end{equation}

In the  't Hooft limit of large $N$ with a fixed ratio $\lambda=N/k$, we need to distinguish two cases. When $Y^{I_1}\neq A$ and $Y_{J_L}^\dagger\neq B$, we get
\begin{equation}
\begin{split}
  \langle O_W \bar O_{W}\rangle \sim & \frac{1}{(N-1)!^2} (N-1)!^2
  \epsilon^{[I]_{N-1}c}_{[J]_{N-1}a}\epsilon^{[K]_{N-1}b}_{[L]_{N-1}c}
  \epsilon^{[L]_{N-1}f}_{[K]_{N-1}d}\epsilon^{[J]_{N-1}e}_{[I]_{N-1}f} \langle W^a_b \bar{W}^d_e\rangle\\
  =  &(N-1)!^4 N \langle\tr (W \bar{W})\rangle\\
  \sim &(N-1)!^4 N^{2L+2}.
\end{split}
\end{equation}
Here we have omitted  the spacetime dependence explicitly  because they can be easily put back at the end of the calculation. When $Y^{I_1}= A$ or $Y_{J_L}^\dagger = B$ the operator factorizes \cite{Berenstein:2003ah,Balasubramanian:2004nb}, so the combinatorics of contractions is
different.
For instance, when $W=A V$ we have
\begin{equation}
O_W=\det A \epsilon^{[K]_{N-1}b}_{[L]_{N-1}c}B^{[L]_{N-1}}_{[K]_{N-1}} V^c_{b},
\end{equation}
and then
\begin{equation}
  \langle O_W \bar O_{{W}}\rangle \sim N!(N-1)!^3 N^{2L}=(N-1)!^4 N^{2L+1}.
\end{equation}
A similar analysis applies to the case when $Y_{J_L}^\dagger\neq B$.
Therefore the mixing between factorizing operators and non-factorizing operators  is suppressed in the large $N$ limit.\footnote{This can be checked at two-loop order by a simple large $N$ counting.}
In this paper we only consider operators with $Y^{I_1}\neq A$ and $Y_{J_L}^\dagger\neq B$.

\subsection{Two-loop open spin-chain Hamiltonian}
We now derive the two-loop anomalous dimension matrix for determinant like operators in the  't Hooft limit.
We need to consider the mixing of two operators
\begin{equation}
W=Y^{I_1}Y_{J_1}^\dagger \cdots Y^{I_L}Y_{J_L}^\dagger,~~~\bar{\tilde{W}} =Y^{M_L}Y_{N_L}^\dagger \cdots Y^{M_1}Y_{N_1}^\dagger
\end{equation}
 where $Y^{I_1}\neq A$, $Y_{N_1}^\dagger\neq \bar{A}$, $Y_{J_L}^\dagger\neq B$ and $Y^{M_L}\neq \bar{B}$.  Keeping one  $A$ and one  $B$ uncontracted with the corresponding $\bar A$ and $\bar B$, we get
\begin{equation}
\begin{split}
  \langle O_W \bar O_{\tilde{W}}\rangle_{\mathrm{2-loop}} \sim &  (N-1)^2
  \epsilon^{[I]_{N-2}ic}_{[J]_{N-2}ja}\epsilon^{[K]_{N-2}kb}_{[L]_{N-2}lc}
  \epsilon^{[L]_{N-2}mf}_{[K]_{N-2}sd}\epsilon^{[J]_{N-2}qe}_{[I]_{N-2}pf}
  \langle A^{j}_iB^{l}_k \bar A^{p}_q \bar B^{s}_m
  W^a_b \bar{\tilde{W}}^d_e\rangle_{\mathrm{2-loop}}\\
=&(N-2)!^2 (N-1)!^2
  \delta^{qe}_{ja}\delta^{ic}_{pf}\delta^{mf}_{lc}\delta^{kb}_{sd}
  \langle A^{j}_iB^{l}_k \bar A^{p}_q \bar B^{s}_m
  W^a_b \bar{\tilde{W}}^d_e\rangle_{\mathrm{2-loop}}.
\end{split}
\end{equation}
Contractions of the generalized Kronecker deltas give
\begin{equation}
\begin{split}\label{WAB}
 &\langle \delta^{qe}_{ja}\delta^{ic}_{pf}\delta^{mf}_{lc}\delta^{kb}_{sd}
   A^{j}_iB^{l}_k \bar A^{p}_q \bar B^{s}_mW^a_b \bar{\tilde{W}}^d_e \rangle_{\mathrm{2-loop}}\\
 =  &(N-2)\langle\tr(W \bar{\tilde{W}})\tr(A \bar A)\tr(B \bar B)
  -\tr( \bar{\tilde{W}} W\bar B B)\tr(A \bar A)-\tr(A \bar A W \bar{\tilde{W}})\tr(B \bar B)+\tr(W \bar B B\bar{\tilde{W}} A \bar A)\rangle_{\mathrm{2-loop}}\\
   &+\langle\tr(W\bar{\tilde{W}})\tr(A B\bar B \bar A)-\tr(W\bar B \bar A A B\bar{\tilde{W}})-\tr(W\bar{\tilde{W}} A B\bar B \bar A)
   +\tr(W\bar B \bar A )\tr(A B\bar{\tilde{W}})\rangle_{\mathrm{2-loop}}.
\end{split}
\end{equation}
One can check that in the large $N$ limit the first, second and third terms in the second line give bulk, right and left boundary contributions respectively,
and the contributions from other terms are suppressed.
For example, one part of the leading contribution from the second term corresponds to the contraction
\begin{equation}
-(N-2)\langle\tr( \bar{\tilde{W}} W\bar B B)\rangle_{\mathrm{connected, \, 2-loop}}\tr(\contraction{}{A}{}{\bar A}
A \bar A)\sim \dfrac{N^{2L+6}}{k^2}.
\end{equation}
Note that the contraction between $A$ and $\bar A$ gives a factor $N^2(N-1)^{-1}$, here the factor $(N-1)^{-1}$ is from avoiding repeatedly counting of contractions.
The  Hamiltonian of the bulk part the open chain is the same as that of the closed spin chain which was derived in \cite{Minahan:2008hf,Bak:2008cp}. We need to consider the boundary contributions.
We first focus on the left boundary corresponding to the term
\begin{equation}
\langle-\tr(A \bar A W  \bar{\tilde{W}})\tr(B \bar B)\rangle_{\mathrm{2-loop}}\rightarrow
\langle-\tr(A \bar A Y^{I_1}Y_{J_1}^\dagger Y^{M_1}Y_{N_1}^\dagger)\rangle_{\mathrm{2-loop}}.
\end{equation}

Contributions from wave function renormalization (self-interactions) are proportional to $\delta^{I_1}_{N_1}$ and thus flavor blind. Because $Y^{I_1}\neq A$ and $Y_{N_1}^\dagger\neq \bar{A}$, contributions from gluon exchange and  fermion exchange are also flavor blind.
We only need to consider contribution from sextet scalar potential $V_B$. Then we get

\begin{equation}
H'_{\mathrm{left}}=\frac{\lambda^2}{2}\left(
\frac{1}{2} \delta^{I_1}_{J_1}\delta^{M_1}_{N_1}+2\delta^{M_1}_{1}\delta^{1}_{J_1}\delta^{I_1}_{N_1}-\delta^{I_1}_{N_1}\delta^{M_1}_{J_1}+C \delta^{I_1}_{N_1}\delta^{M_1}_{J_1}
  \right).
\end{equation}
Here the normalization is fixed by comparing with bulk Hamiltonian from sextet scalar potential.  The constant $C$ comes from the contributions from gluon exchange, fermion exchange and self-interactions. An analogous discussion  applies to the right boundary.
We will show in Appendix~\ref{bps} that the anomalous dimension of the operator with $W=(A_2B_2)^L$ is zero in the large $N$ limit, which allows us to  determine the sum of the constant $C$ and a similar constant from the right boundary.
At the end the total Hamiltonian is given by
\begin{equation}\label{Hamiltonian}
\begin{split}
H=& \lambda^2\sum_{l=2}^{2L-3}\left(
\mathbb{I}-\mathbb{P}_{l,l+2}+\frac{1}{2}\mathbb{P}_{l,l+2}\mathbb{K}_{l,l+1}+\frac{1}{2}\mathbb{P}_{l,l+2}\mathbb{K}_{l+1,l+2}
\right) Q_1^AQ_{2L}^B\\
  &+\lambda^2Q_1^A\left(
  \mathbb{I}+\frac{1}{2}\mathbb{K}_{1,2}-\mathbb{P}_{1,3}+\frac{1}{2}\mathbb{P}_{1,3}\mathbb{K}_{1,2}+\frac{1}{2}\mathbb{P}_{1,3}\mathbb{K}_{2,3}
  \right)Q_1^AQ_{2L}^B\\
&+\lambda^2Q_{2L}^B\left(
\mathbb{I}+\frac{1}{2}\mathbb{K}_{2L-1,2L}-\mathbb{P}_{2L-2,2L}+\frac{1}{2}\mathbb{P}_{2L-2,2L}\mathbb{K}_{2L-2,2L-1}+\frac{1}{2}\mathbb{P}_{2L-2,2L}\mathbb{K}_{2L-1,2L}
\right)
\\
&Q_1^AQ_{2L}^B+\lambda^2(\mathbb{I}-Q_2^{\bar A})Q_1^AQ_{2L}^B+\lambda^2(\mathbb{I}-Q_{2L-1}^{\bar B})Q_1^AQ_{2L}^B
\end{split}
\end{equation}
where the  trace operator $\mathbb{K}$ and permutation operator $\mathbb{P}$ are   defined as
\begin{equation}
(\mathbb{K}_{ij})^{I_iI_j}_{J_iJ_j}=\delta^{I_iI_j}\delta_{J_iJ_j},~~~
(\mathbb{P}_{ij})^{I_iI_j}_{J_iJ_j}=\delta^{I_j}_{J_i}\delta_{J_i}^{I_j},
\end{equation}
and the $Q$ operators are defined as \cite{Berenstein:2005vf}
\begin{equation}
Q^{\phi}|\phi\rangle=0,~~~Q^{\phi}|\psi\rangle=|\psi\rangle,~~~\mathrm{for }~\psi \neq \phi.
\end{equation}
Half of the $\frac{1}{2} \mathbb{K}_{1,2}$ ($\frac{1}{2}\mathbb{K}_{2L-1,2L}$)  term in  (\ref{Hamiltonian}) comes from the third (second) term in (\ref{WAB}),
and another half comes from the first term in (\ref{WAB}).

\section{Integrability from coordinate Bethe ansatz}
In this section we discuss the integrability of the above open spin chain in the framework of coordinate Bethe ansatz.
The reflection equations are  necessary conditions for the integrability of the open spin chain Hamiltonian.
We want to know whether the boundary reflection matrices satisfy the reflection equations or not.

The vacuum of this open chain is chosen to be
 \be W= (A_2B_2) \cdots (A_2B_2).\ee

The one-particle excitations include
\bea
\mbox{bulk odd site}&& (A_2B_2) \cdots (A_1 B_2)\cdots (A_2B_2) \\
                     && (A_2B_2)\cdots (B_1^\dag B_2)\cdots (A_2B_2)\\
\mbox{bulk even site}&& (A_2B_2) \cdots (A_2 B_1)\cdots (A_2B_2) \\
                     && (A_2B_2)\cdots (A_2A_1^\dag)\cdots (A_2B_2)\\
\mbox{left boundary}&& (B_1^\dag B_2)\cdots (A_2 B_2)\\
\mbox{right boundary}&& (A_2 B_2)\cdots (A_2 A_1^\dag).
\eea
We denote the open chain as $(1)(2)\cdots(x)\cdots(L)$ with every site $(x)$ containing two fields.  Then the above excitations can be simply denoted
as \bea && |x\rangle_{A_1}, 2\le x\le L, \nn\\
        && |x\rangle_{B_1^\dag}, 1\le x\le L, \nn\\
        && |x\rangle_{B_1}, 1\le x\le L-1, \nn\\
        && |x\rangle_{A_1^\dag}, 1\le x\le L,
\eea
where $|1\rangle_{B_1^\dag}$ is the left boundary excitation state and  $|L\rangle_{A_1^\dag}$ is the right boundary excitation state while all others are bulk one-particle excitation state.

Let us begin with \be|k\rangle_{B^\dag_1}=\sum_{x=1}^{L}f_{B_1^\dag}(x)|x\rangle_{B_1^\dag}, \ee
where \be f_{B_1^\dag}(x)=F_{B_1^\dag}e^{ikx}+\tilde{F}_{B_1^\dag}e^{-ikx}. \ee

On the states $|x\rangle_{B_1^\dag}$, the Hamiltonian acts as follows
\be H|x\rangle_{B_1^\dag}=\lambda^2(2|x\rangle_{B_1^\dag}-|x+1\rangle_{B_1^\dag}-|x-1\rangle_{B_1^\dag}), \ee
when $2\le x\le L-1$, and
\be H|1\rangle_{B_1^\dag}=\lambda^2(|1\rangle_{B_1^\dag}-|2\rangle_{B_1^\dag}), \ee
\be H|L\rangle_{B_1^\dag}=\lambda^2(2|L\rangle_{B_1^\dag}-|L-1\rangle_{B_1^\dag}). \ee
So we get \bea H|k\rangle_{B^\dag_1}&=&\lambda^2\sum_{x=2}^{L-2}(2f_{B_1^\dag}(x)-f_{B_1^\dag}(x-1)-f_{B_1^\dag}(x+1))|x\rangle_{B_1^\dag}\nonumber\\
&+&\lambda^2(f_{B_1^\dag}(1)-f_{B_1^\dag}(2))|1\rangle_{B_1^\dag}+\lambda^2(2f_{B_1^\dag}(L)-f_{B_1^\dag}(L-1))|L\rangle_{B_1^\dag}. \eea
Then equation \be  H|k\rangle_{B_1^\dag}=E(k)|k\rangle_{B_1^\dag},\ee leads to  the following dispersion relation
\be E(k)=\lambda^2(2-2\cos k),\ee
and \bea f_{B_1^\dag}(1)&=&f_{B_1^\dag}(0),\label{barb1left}\\
f_{B_1^\dag}(L+1)&=&0.\label{barb1right}\eea
Since the reflections of $B_1^\dag$ excitation at both sides are diagonal, we define the left reflection coefficient to be
\be K_{L,\, B_1^\dag}=F_{B_1^\dag}/\tilde{F}_{B_1^\dag}, \ee
and the right reflection coefficient to be\footnote{We have taken into account that for every excitation, there are  $L-1$ bulk sites.} \be K_{R,\, B_1^\dag}=e^{2ik(L-1)}F_{B_1^\dag}/\tilde{F}_{B_1^\dag}. \ee
They are determined by eqs.~(\ref{barb1left}) and (\ref{barb1right}), respectively. The results are
\bea K_{L, \, B_1^\dag}&=&e^{-ik},\\
K_{R,\, B_1^\dag}&=&-e^{-4ik}.\eea

For the other three excitations, the computations are similar. So we only list the action of the Hamiltonian,
obtained boundary conditions and reflection coefficients.
For $|x\rangle_{A_1}, \, 2\le x\le L$ we have
\bea
H|x\rangle_{A_1}&=&\lambda^2(2|x\rangle_{A_1}-|x+1\rangle_{A_1}-|x-1\rangle_{A_1}),~~3\le x\le L-1\\
 H|2\rangle_{A_1}&=&\lambda^2(2|2\rangle_{A_1}-|3\rangle_{A_1}),\\
H|L\rangle_{A_1}&=&\lambda^2(|L\rangle_{A_1}-|L-1\rangle_{A_1}).\eea

This gives \be f_{A_1}(1)=0, \, f_{A_1}(L+1)=f_{A_1}(L), \ee
which leads to \be K_{L,\, A_1}=-e^{-2ik}, K_{R,\, A_1}=e^{-3ik}. \ee

For $|x\rangle_{B_1}, \, 1\le x\le L-1$, we have \bea
H|x\rangle_{B_1}&=&\lambda^2(2|x\rangle_{B_1}-|x+1\rangle_{B_1}-|x-1\rangle_{B_1}),~~2\le x\le L-2\\ H|1\rangle_{B_1}&=&\lambda^2(|1\rangle_{B_1}-|2\rangle_{B_1}),\\
H|L-1\rangle_{B_1}&=&\lambda^2(2|L-1\rangle_{B_1}-|L-2\rangle_{B_1}).\eea
this leads to\be f_{B_1}(1)=f_{B_1}(0), \, f_{B_1}(L)=0, \ee
then \be K_{L,\, B_1}=e^{-ik}, \, K_{R,\, B_1}=-e^{-2ik}. \ee

Finally for $|x\rangle_{A_1^\dag}, \, 1\le x\le L$, we have
\bea
H|x\rangle_{A_1^\dag}&=&\lambda^2(2|x\rangle_{A_1^\dag}-|x+1\rangle_{A_1^\dag}-|x-1\rangle_{A_1^\dag}),~~2\le x\le L-1\\
H|1\rangle_{A_1^\dag}&=&\lambda^2(2|1\rangle_{A_1^\dag}-|2\rangle_{A_1^\dag}), \\
H|L\rangle_{A_1^\dag}&=&\lambda^2(|L\rangle_{A_1^\dag}-|L-1\rangle_{A_1^\dag}). \eea
This gives \be f_{A_1^\dag}(0)=0,\, f_{A_1^\dag}(L)=f_{A_1^\dag}(L+1), \ee
and \be K_{L,\, A_1^\dag}=-1, \, K_{R, \,A_1^\dag}=e^{-3ik}.\ee
With the order of the excitations  as $A_1, B_1^\dag, A_1^\dag, B_1$, the left reflection matrix is
\bea  K_L=\left(\begin{array}{cccc}
                  -e^{-2ik} &  & &  \\
                   & e^{-ik} &  &  \\
                   &  & -1 &  \\
                   & &  & e^{-ik}
                \end{array}\right), \eea

and the right reflection matrix is
\bea  K_R=\left(\begin{array}{cccc}
                  e^{-3ik} &  & &  \\
                   & -e^{-4ik} &  &  \\
                   &  & e^{-3ik} &  \\
                   & &  & -e^{-2ik}
                \end{array}\right). \eea

The two reflection matrices are diagonal in the chosen natural basis. This is quite different from the results in \cite{Bai:2017jpe}, where
the reflection matrices are anti-diagonal in the same basis\footnote{A non-supersymmetric flavored ABJM theory was constructed in \cite{Wu:2017tkc}, where the corresponding reflection matrices are diagonal.}. Also notice that each excitation always has Dirichlet boundary condition on one end of the
open chain, and Neumann boundary condition on the other end. This is different from the SYM case \cite{Balasubramanian:2004nb,Berenstein:2005vf} where the boundary conditions are always left-right symmetric.
The S-matrix in ABJM theory can be found in \cite{Ahn:2009zg}. It satisfies the Yang-Baxter equation
\begin{equation}
S_{12}(k_1,k_2)S_{13}(k_1,k_3)S_{23}(k_2,k_3)=S_{23}(k_2,k_3)S_{13}(k_1,k_3)S_{12}(k_1,k_2).
\end{equation}
Now we are ready to check the reflection equations.
It can be straightforward to verify that reflection equations are satisfied
\begin{align}
&K_{L2}(k_2)S_{12}(k_1,-k_2)K_{L1}(k_1)S_{21}(-k_2,-k_1)=S_{12}(k_1,k_2)K_{L1}(k_1)S_{21}(k_2,k_1)K_{L2}(k_2),\\
&K_{R2}(-k_2)S_{21}(k_2,-k_1)K_{R1}(-k_1)S_{12}(k_1,k_2)=S_{21}(-k_2,-k_1)K_{R2}(-k_1)S_{12}(k_1,-k_2)K_{R2}(-k_2).
\end{align}

The $\frac{1}{2} \mathbb{K}_{1,2}$ and $\frac{1}{2}\mathbb{K}_{2L-1,2L}$ terms in the Hamiltonian (\ref{Hamiltonian}) have no effect in the above calculation.
To understand their role in the coordinate Bethe ansatz, one needs to consider impurities $A_2^\dag$ and $B_2^\dag$.
These impurities can be described as bound states of the form $\phi \phi^\dag, \phi=A_1, B_1^\dagger$.
Although not shown here, we have checked that the $\frac{1}{2} \mathbb{K}_{1,2}$ and $\frac{1}{2} \mathbb{K}_{2L-1,2L}$ terms in the Hamiltonian are necessary in the construction of the  eigenstates involving  $\phi \phi^\dag$ scattering and the above bound states using coordinate Bethe ansatz.

\section{Conclusions and discussions}
We have obtained the two-loop Hamiltonian of the open spin chain corresponding to the determinant like operators in ABJM theory which are dual to open strings attached to D4-branes wrapping cycles in $\mathbb{CP}^3$. The Hamiltonian is different from the periodic spin chain only in the boundary terms. Using  the coordinate Bethe ansatz, we  present strong evidence that the  Hamiltonian may be integrable. In other words, the giant graviton may provide integrable boundary conditions for the open string. It is possible to go beyond the two loop order to an all loop prediction which is  similar to previous studies in the SYM context \cite{Hofman:2007xp,Ahn:2008df} using symmetries as the guide, and could even further to solve the full open string spectrum through boundary thermodynamical Bethe ansatz and/or Y-system which have already been done in the SYM case \cite{Bajnok:2012xc,Bajnok:2013wsa}. To have a more solid ground for integrability of our two loop Hamiltonian, it would be better to have an algebraic Bethe ansatz construction \cite{BCOW} as people have done in the SYM theory \cite{Nepomechie:2011nz}.

\section*{Acknowledgments}
We would like to thank Nan Bai, Yunfeng Jiang for very helpful  discussions.
The work of J.-B.~W.  was supported  by the National Natural Science Foundation of China, Grant No.\ 11575202.
The work of H.~O. was supported by the grant ``Exact Results in Gauge and String Theories'' from the Knut and Alice Wallenberg foundation.
H.~O. and H.-H.~C. gratefully acknowledge the kind hospitality of Tianjin University during the course of this work.

\appendix

\section{ Vacuum of the open chain}\label{bps}

In this appendix, we show that  the anomalous dimension of the operator
\begin{equation}
O_0 = \epsilon_{a_1...a_N}\epsilon^{b_1...b_N}(A_1 B_1)^{a_1}_{b_1}...(A_1 B_1)^{a_{N-1}}_{b_{N-1}}\big((A_2 B_2)^L\big)^{a_N}_{b_N}
\end{equation}
is  suppressed in the large $N$ limit with $\lambda=N/k$ fixed. As discussed in \cite{Kristjansen:2008ib}, at two-loop order the contribution from bosonic D-terms, gluon exchange, fermion exchange from fermionic D-terms and self-interactions cancel for operators in the $SU(2)\times SU(2)$ sector, and the fermionic F-terms do not contribute to the anomalous dimension. We only need to consider the contributions from bosonic F-terms \cite{Benna:2008zy}
\begin{equation}
\begin{split}
V_F^{\mathrm{bos}}=-\dfrac{16 \pi^2}{k^2}\tr(&A^{\dag i}B^{\dag }_jA^{\dag k}A_iB^jA_k-A^{\dag i}B^{\dag }_jA^{\dag k}A_kB^jA_i\\
&+B^{\dag }_iA^{\dag j}B^{\dag }_kB^iA_jB^k-B^{\dag }_iA^{\dag j}B^{\dag }_kB^kA_jB^i).
\end{split}
\end{equation}
Using (\ref{WAB}) one can check that the anomalous dimension of $O_0$ is subleading in $1/N$.


\providecommand{\href}[2]{#2}\begingroup\raggedright\endgroup

\end{document}